\documentclass[12pt]{article}
\usepackage{amssymb,latexsym}
\usepackage[mathscr]{eucal}
\usepackage{amsmath,amssymb}
\usepackage{latexsym}
\usepackage{bbm}
\usepackage{bm}

\newcommand{\beq}{\begin{equation}}
\newcommand{\eeq}{\end{equation}}
\newcommand{\beqn}{\begin{eqnarray}}
\newcommand{\eeqn}{\end{eqnarray}}

\newcommand{\nn}{\nonumber}

\numberwithin{equation}{section}

\newcommand{\be}{\begin{equation}}
\newcommand{\ee}{\end{equation}}
\newcommand{\ba}{\begin{eqnarray}}
\newcommand{\ea}{\end{eqnarray}}
\newcommand{\bdm}{\begin{displaymath}}
\newcommand{\edm}{\end{displaymath}}

\def\a{\alpha}

\newcommand{\ie}{{\it i.e.\ }}

\DeclareMathAlphabet{\mathpzc}{OT1}{pzc}{m}{it}
\def\bea{\begin{eqnarray}}
\def\eea{\end{eqnarray}}
\def\beas{\begin{eqnarray*}}
\def\eeas{\end{eqnarray*}}
\def\sla{\raise.15ex\hbox{$/$}\kern-.57em}

\def\bea{\begin{eqnarray}}
\def\eea{\end{eqnarray}}
\def\de{\partial}

\def\sla{\raise.15ex\hbox{$/$}\kern-.57em}
\def\ie{{\it i.e.}~}

\def\a{\alpha}

\def\cA{{\cal A}}

\def\cE{{\cal E}}

\def\cI{{\cal I}}

\def\cK{{\cal K}}

\def\cM{{\cal M}}
\def\cN{{\cal N}}

\def\cT{{\cal T}}

\begin{document}
\begin{titlepage}
\begin{flushright}
CERN-PH-TH/256 \\
{ROM2F/2007/22}\\
UCLA/07/TEP/29
\end{flushright}
\begin{center}
{\large \sc  Enriques and Octonionic  \\ Magic Supergravity Models }\\
\vspace{1.0cm}
{\bf Massimo Bianchi$^{1,2}$} and {\bf Sergio Ferrara$^{1,3,4}$}\\
$^1${\sl Physics Department, Theory Unit, CERN \\ CH 1211, Geneva
23,
Switzerland }\\
$^2${\sl Dipartimento di Fisica, Universit\'a di Roma ``Tor Vergata''\\
 I.N.F.N. Sezione di Roma ``Tor Vergata''\\
Via della Ricerca Scientifica, 00133 Roma, Italy}\\
$^3${\sl INFN - Laboratori Nazionali di Frascati \\ Via Enrico
Fermi 40, 00044 Frascati, Italy}\\
$^4${\sl Department of Physics and Astronomy \\
University of California, Los Angeles, CA USA}
\\
\end{center}
\vskip 2.0cm
\begin{center}
{\large \bf Abstract}
\end{center}

We  reconsider the Enriques Calabi Yau (FHSV) model and its string
derivation and argue that the Octonionic magic supergravity theory
admits a string interpretation closely related to the Enriques
model. The uplift to $D=6$ of the Octonionic magic model has 16
abelian vectors related to the rank of Type I and Heterotic
strings.

\vfill
\end{titlepage}

\section{Introduction}

Among the magical supergravities \cite{GST}, related to the famous
magic square of Freudenthal, Rozenfeld and Tits of the division
algebras $R,C,H,O$, there is just one, the Octonionic model, which
cannot be obtained as a truncation of ${\cN}=8$ supergravity. This
is obviously due to the fact that such theory is based on the real
form $E_{7(-25)}$ of the exceptional group $E_7$ while ${\cN}=8$
supergravity is based on the real form $E_{7(+7)}$. As a
consequence, the corresponding moduli space of the ${\cN}=2$ and
${\cN}=8$ supergravities based on $E_{7(-25)}$ (Octonions) and
$E_{7(+7)}$ (split Octonions) are very different. The former is
the rank 3, 54-dimensional K\"ahler space $E_{7(-25)}/E_{6}\times
U(1)$. The latter is the rank 7, 70-dimensional non-K\"ahler space
$E_{7(7)}/SU(8)$. However, following Gunaydin et al. \cite{GST}
the ${\cN}=2$ $E_{7(-25)}$ model is completely unified since the
28 vectors (27 matter vectors and one graviphoton) mix under
$E_{7(-25)}$ electric-magnetic duality rotations. This symmetry is
not a symmetry of the Lagrangian but only of the field equations.
The maximal symmetry of the Lagrangian, which does not mix
electric and magnetic potentials, being $SU^*(8)$ \cite{Hull}. the
possible relation with the FHSV model \cite{FHSV} comes from the
observation that $E_{7(-25)}$ (unlike $E_{7(7)}$!) contains as
maximal subgroup $SO(2,10)\times SU(1,1)$ and indeed, for the FHSV
model, the space \be {SO(2,10)\over SO(2)\times SO(10)}\times
{SU(1,1)\over U(1)} \ee is the moduli space of complex structure
deformations of the underlying space which is a torus fibration of
an Enriques surface $CY_{FHSV} \approx \cE \times T^2$ with
holonomy $SU(2)\times Z_2$. Indeed we have \be {SO(2,10)\over
SO(2)\times SO(10)}\times {SU(1,1)\over U(1)} \subset
{E_{7(-25)}\over E_{6}\times U(1)} \ee Note that this moduli space
cannot be obtained as a truncation of $E_{7(+7)}/SU(8)$, the
moduli space of ${\cN}=8$ supergravity.

In an analogous way the hypermultiplet moduli space, that includes
deformations of the K\"ahler structure of $CY_{FHSV} \approx \cE
\times T^2$, is obtained by c-map \cite{CFG} to be \be
{SO(12,4)\over SO(12)\times SO(4)} \ee and is a quaternionic
subspace of the exceptional quaternionic manifold obtained by
c-map from the Octonionic magic model \cite{CFG} \be
{SO(12,4)\over SO(12)\times SO(4)} \subset {E_{8(-24)}\over
E_{7}\times SU(2)} \ee

Note that the moduli space cosets of the Octonionic model and of
the FHSV model have the same rank (respectively 3 and 4 for the
special and quaternionic manifolds). The number of vector
multiplets as well as hypermultiplets is augmented by 16 each with
respect to the FHSV model \be n_V^{\bf O} = 27 = 11 +16 \quad ,
\quad n_H^{\bf O} = 28 = 12 +16 \ee and quite remarkably 16 is the
rank of the gauge group in Type I and Heterotic models in $D=10$.

Both models correspond to self-mirror CY threefolds with $h_{11} =
h_{21}= 11$ and $h_{11} = h_{21}= 27$, respectively, and admit an
uplift to $D=6$ with $\cN = (1,0)$ supersymmetry. The $D=6$
interpretation is in terms of $n_T =9$ tensor multiplets, $n_H=12$
hypermultiplets and $n_V=0$ vector multiplets for the FHSV model
and $n_T =9$ tensor multiplets, $n_H=28$ hypermultiplets and
$n_V=16$ vector multiplets for the Octonionic model.

We will give a simple construction of the two models and show
analogies and differences. Electric-magnetic duality in $D=4$ and
special geometry are discussed in Sect.~2. In Sect.~3, we discuss
BPS and non BPS black-holes solutions and attractors in the two
magic models. In Sect.~4 we describe the embedding of the parent
$D=6$ models in Type I superstring and F-theory on Voisin-Borcea
(VB) orbifolds. Our concluding remarks and comments on other magic
models are in Sect.~5. An appendix contains details of the
construction of the Type I superstring models underlying the two
magic models.

\section{Duality rotations}

In this section we discuss the duality  properties \cite{GZ} of
the effective $\cN = 2$ theory for the Enriques CY and the
Octonionic magic model. As we have seen before, there is a common
sector of the two models which comes from the $\cN = (1,0)$ tensor
multiplets after Kaluza Klein reduction from $ D=6 $. This gives
rise to the vector multiplet  (from the tensor multiplets plus KK
vectors) moduli space\footnote{This is the space L(8,0)=L(0,8) in
the notation of \cite{DVV}, while the Octonionic model is L(8,1).}
\be {SU(1,1)\over U(1)}\times {SO(2,10)\over SO(2)\times SO(10)}
\ee The cubic holomorphic polynomial,  from which  the $\cN = 2$
prepotential of the underlying special geometry for the FHSV model
arises, is \be F_{FHSV}(X) = { s \over 2} \eta_{IJ} x^I x^J \ee
where $\eta_{IJ} x^I x^J = x_{10}^2 - \sum_{i=1}^9 x_i^2$. Note
that $F(X)$ has manifest $SO(1,9)$ invariance because of the
Lorentzian contraction of the 10 $X$ coordinates. These originate
from the nine tensor scalars related to the classical moduli space
of those K3 moduli which survive on the Enriques surface.
Therefore $SO(1,9)$ does not give electric-magnetic duality
transformations since it does not mix the electric vectors with
their duals. The moduli corresponding to $ReX$   are the axions
that come from the two forms and have an associated
shift-symmetry. The larger symmetry  $SO(2,10)$ does mix electric
and magnetic field strengths, contrary to the Heterotic string
where an analogous symmetry, $SO(2,n_V)$ T-duality, does not act
as electric-magnetic duality rotations.

Let us now move to the Octonionic model with $h_{11}  =h_{11} =
27$. In this case the cubic polynomial, as it would come      from
a six-dimensional interpretation, is \cite{FMS, AFL} \be F_{OM}(X)
= { s\over 2} ( \eta_{IJ}x^I x^J) - x_I C^I_{ab}v^a v^b \ee where
$\eta_{IJ}=(1,-1,-1,...,-1)$  and $v^a$ ($a =1, ...16$) are the
complex scalars in the $\cN =2$ vector multiplets, that can be
identified with the 6-D vector fields of the Cartan subalgebra
along the two compactified directions.  The structure constants $
C^I_{ab}$ determine the coupling of (tensor multiplet) scalars to
the 16 vectors in $D=6$ in the Cartan subalgebra of $U(16)$. Note
that $ C^I_{ab}$ satisfy the cocycle condition
\cite{AFL,FRS,RS,NS} \be \eta_{IJ}C^I_{(ab} C^J_{cd)}=0 \ee which
follows from gauge invariance of the six dimensional theory in the
Coulomb   phase where $U(16)$ is Higgsed  to $U(1)^{16}$ and
massive   states are integrated out. If we demand that each choice
in the complex structure of the Enriques surface be
$SO(1,9)/SO(9)$ equivalent, then  $ C^I_{ab} $   must be the
(symmetric) $\gamma$ matrices of the $SO(1,9)$ Clifford algebra,
being the 16 vectors a chiral $ SO(1,9)$ spinor representation,
which is real and inequivalent to $16'$. Still, the $SO(1,9)$
symmetry is not an e.m. duality in 4D since it does not mix
electric with magnetic field strengths. However when
$SO(1,9)\rightarrow SO(2,10)$ an $SO(2,10)$ chiral spinor
representation has  32 real components and indeed in this case the
action of $SO(2,10)$ mixes the 4D vectors with their duals. This
phenomenon is similar to the action of T-duality on R-R fields in
type  II superstring theory\footnote{Indeed $SO(6,6)$ and
$SO(2,10)$ are two inequivalent real forms of $SO(12)$.}. We also
remark that the requirement that each point on the $SO(1,9)/ SO(9)
$ moduli space gives equivalent Physics is the key to the
enlargement of the manifest $ SO(2,10)\times SU(1,1)$ to the
exceptional group $E_{7(-25)}$. Indeed,  under $ SO(2,10)$ the
$12+16=28$ vectors, together with their duals, form a 56
dimensional space as follows $56 = (12,2)+(32,1)$. This is
identical to the decomposition which takes place in type II
supergravity if one  decomposes $E_{7(7)}$ with respect to the
T-duality  sub group $SO(6,6)$ and the axion-dilaton symmetry
$SL(2,R)$ \be 56=(12_{NS-NS}, 2) + (32_{R-R}, 1) \ee

The above consideration explains our previous remark.

\section{Extreme Black-Holes and Attractors}

The two models under consideration have also interesting
properties as far as extreme black-hole solutions are concerned.
Since their moduli spaces fall in the classification of symmetric
spaces in the literature \cite{BFGM}, we just comment on their
attractor solutions. Both models have both BPS and  non BPS
black-holes depending on which orbit the charge vector (24
dimensional in the first case, 56 dimensional in the second case)
lies in. The classification of orbits for the FHSV model yields
\cite{BFGM} \bea
BPS  \qquad &&{SU(1,1) \times  SO(2,10) \over SO(2) \times
SO(10)} \nn \\
NBPS \ (Z\neq 0)  \qquad &&{SU(1,1) \times  SO(2,10) \over SO(1,1)
\times
SO(1,9)} \nn \\
NBPS \ (Z=0) \qquad &&{SU(1,1) \times  SO(2,10) \over SO(2) \times
SO(2,8)} \eea The 11 complex moduli  are  all fixed   in the BPS
orbit while   there is a        moduli space in the NBPS case
\cite{FM} (the $\cN =2$ central charge $Z$ is a section of the
K\"ahler $U(1)$ bundle) \bea
NBPS \ (Z\neq 0)  \quad &&SO(1,9)/SO(9) \nn \\
NBPS \ (Z=0)  \quad &&SO(2,8)/SO(2)\times SO(8) \eea The previous
considerations exhaust the analysis of the FHVS model.

For  the Octonionic  theory   the classification of attractors is
as follows, the charge orbits         are \cite{BFGM} \bea
BPS \  \qquad &&{ E_{7(-25)} \over E_6}\nn \\
NBPS \ (Z\neq 0)  \qquad &&{ E_{7(-25)} \over E_{6(-26)}}\nn \\
NBPS\ (Z=0) \qquad &&{ E_{7(-25)} \over E_{6(-14)}}\eea

The  residual moduli space of the non BPS  attractors are
\cite{FM}
\bea
NBPS\ (Z\neq 0)  \qquad &&{ E_{6(-26)} \over F_4}\nn \\
NBPS\  (Z=0)  \qquad &&{ E_{6(-14)} \over SO  (2) \times  SO(10)}
\eea Note that all moduli spaces of all non BPS orbits of the FHVS
and Octonionic magic model have in common the  restricted moduli
space $SO(1,8)/SO(8)$. This is the tensor multiplet moduli space
of non BPS self-dual string for 9 tensor multiplets, one of the
tensor moduli being fixed by the six-dimensional version of the
attractor mechanism \cite{ADFL, FG, AFMT}.

For all these models, the classical 4D black-hole entropy of the
attractor solutions is given by the following formula \cite{FG,
FGK} \be S = \pi \sqrt{|\cI_4|} \ee where $ \cI_4$ is an
electric-magnetic duality invariant combination of the electric
and magnetic charges of the theory. For the FHSV model $ \cI_4$ is
the unique singlet in the product of four $(2,12)$ irreps of
$SL(2)\times SO(2,10)$. For the Octonionic model $ \cI_4$  is the
unique singlet in the product of four 56 irreps of $E_{7(-25)}$.

\section{Model Building}

As previously observed, the two $\cN =2$ supergravity models can
be obtained from compactification on $T^2$ of $\cN =(1,0)$ chiral
supergravity models in $D=6$. Both of them have the same number of
tensor multiplets, $n_T=9$. While the parent of the FHSV has $n_H
=12 $ hypermultiplets and $n_V=0$ vector multiplets, the parent of
the Octonionic magic model has $n_H =28 $ and $n_V=16$.

Models of this kind can be embedded in string theory. The most
efficient way is to consider unoriented descendants of Type IIB
superstrings on $K3$ \cite{MBAS1, MBAS2} which in many cases can
be related to F-theory compactifications on elliptically fibered
CY spaces \cite{CVFt, MV} with constant dilaton \cite{ASFt}.
Perturbative Heterotic models can only have one tensor multiplet
in their massless spectrum and are thus unsuitable for our
purposes\footnote{Including NS5-branes may lead to models with
several tensor multiplets that however lack a full-fledged string
description. In some cases these models can be related to M-theory
compactifications on $K3 \times S^1/Z_2$ with M5-branes,
supporting tensor multiplets}. For certain choices of the
compactification, including the choice of the internal gauge
bundle, heterotic $\cN =(1,0)$ models in $D=6$ can be related to
F-theory or to Type I. In particular the compactification on
$T^4/Z_2$ with gauge group $U(16)$ \cite{BPSSW} can be shown to be
dual to a Type I compactification found in \cite{MBAS2} and
recently discussed in \cite{MBJFMinst} as a playground for
non-perturbative effects.

Focussing on unoriented descendants of the Type IIB superstring in
$D=6$ \cite{MBAS1, MBAS2, ABPSS, GP, GJ, DP, JP, BZ, ZK}, one
starts with $\cN=(2,0)$ supergravity coupled to 21 tensor
multiplets, each containing an anti self-dual tensor and 5
scalars. The moduli space is $SO(5,21)/SO(5)\times SO(21)$. The
unoriented worldsheet parity projection of the closed string
spectrum, coded in the Klein bottle amplitude, produces
$\cN=(1,0)$ supergravity coupled to $n^{cl}_T$ tensor multiplets
and $n^{cl}_H$ `neutral' hypermultiplets. Since both kinds of
matter multiplets descend from the 21 $\cN=(2,0)$ tensor
multiplets, of the parent Type IIB theory, one has a constraint
\be n^{cl}_T + n^{cl}_H = 21 \ee Explicit constructions have
produced models with $n^{cl}_T$ ranging from 0 \cite{ABPSS} to 19
\cite{CVFt, MV, ASFt, BZ, ZK}. Since the open string spectrum
cannot produce any massless tensor multiplet, it is impossible to
exceed $n_T=19$ in this kind of models.

Chiral anomaly cancellation in $D=6$ \cite{GSW}, which is
equivalent to R-R tadpole cancellation in this kind of
compactifications \cite{MBJFM, IRU}, puts severe constraints on
the massless spectrum. In particular absence of gravitational
anomalies requires \be 29 n_T + n_H - n_V = 273
\label{gravanom}\ee Notice that (anti) self-dual antisymmetric
tensor do contribute to the anomaly. Tensorini and hyperini have,
say, L chirality while gaugini and gravitini have R chirality and
contribute with the opposite sign to the anomaly. One can
immediately conclude that $n_T = 9$ is the maximum value for a
theory without vector multiplets and indeed a theory with $n_V=0$,
$n_T=9$ and $n_H=12$ is completely free from anomaly in the sense
that the anomaly polynomial is exactly zero as for the parent Type
IIB theory. In fact the latter is exactly twice the former (2
gravitini, $2\times 21$ L fermions and $5-21$ tensors). As we will
see, this model corresponds to a Type I theory without open
strings \cite{DP,ABPSS} or to F-theory on a Voisin - Borcea (VB)
orbifold that makes use of the freely-acting Enriques involution
of $K3$.

When vector multiplets are present, the irreducible gauge anomaly
is proportional to the quartic Casimir and cancels only for a very
restricted class of models, \ie the ones for which \be Tr_{Adj}
F^4 = \sum_H Tr_{R_H} F^4 \label{guageanom}\ee In Type I models
vector and charged hypermultiplets correspond to open string
excitations of various D-branes present in the background and
coded in the Annulus amplitude and its M\"obius strip projection.
In perturbative models, open strings have only two ends and they
can at most tranform in the product of two fundamental
representations of classical groups.

Once (\ref{guageanom}) (\ref{gravanom}) are  satisfied, one can
invoke a generalization of the G-S mechanism to cancel the
left-over reducible gauge, gravitational and mixed anomalies
\cite{GS, GSW, ASGS}. Actually there are two kinds of mechanisms.
The first one involves (anti) self-dual antisymmetric tensors and
serves to cancel anomalies of the form \be I_{4+4} = {1\over 2}
\sum_I X^I_4 \wedge X^I_4 \ee where $I=0,1,...n_T$ for the
mechanism to work at all. The second one involves 4-forms dual to
axions \cite{BPSSW, Anas} and serves to cancel anomalies of the
form\footnote{Four-dimensional remnants of anomaly cancellation
are the generalized Chern-Simons couplings discussed in \cite{GCS,
AKR, ABDK,VPR}.} \be I_{2+6} = \sum_h X^h_2 \wedge X^h_6 \ee where
$h=1,...n_H$ for the mechanism to work at all. In string theory
modular invariance and tadpole cancellation guarantee the
necessary couplings \cite{MBJFM} \be L_{GSS} = \sum_I C_2^I \wedge
X_4^I + \sum_h [ C_4^h \wedge X_2^h + C_0^h X_6^h ] \ee Terms of
the form $C_4 \wedge X_2 \equiv C_4 \wedge TrF$ can be dualized to
$*dC_0\wedge A = A^\mu \de_\mu C_0$. The field $C_0$ is a
St\"uckelberg field for $A$ or in other words the (abelian) gauge
field $A$ gauges the axionic shift symmetry and becomes massive.
The mechanism can take place in a supersymmetric fashion and lifts
entire hypermultiplets.

In relation to gauge anomaly cancellation, the antisymmetric
combinations of two 8-dimensional representations such as the 28
of $SO(8)$ (Adjoint) or the 28 and $28^*$ of $U(8)$ as well as the
27 of $Sp(8)$ (the antisymmetric singlet decouples) play a
peculiar role. Their contribution to the irreducible anomaly
vanishes and they can thus appear in arbitrary number in the
spectrum. Indeed in F-theory compactifications on VB orbifolds,
that we will momentarily review briefly, the singularities of the
fibration are of $D_4$ type and give rise to products of $SO(8)$
gauge groups and hypers in the Adjoint.

Before doing that, let us briefly recall some aspects of the
low-energy effective action which are relevant for our analysis.
First of all $\cN=(1,0)$ supersymmetry in $D=6$, very much like
$\cN=2$ supersymmetry in $D=4$, prevents neutral coupling of
hypers to vectors. As a consequence the gauge coupling can only
depend on the real scalars in the tensor multiplets \cite{NS, FMS,
FRS, RS}. For perturbative heterotic string compactifications, the
only such scalar is the dilaton and the dependence is linear (tree
level) plus a constant (one loop GS counterterm). In Type I models
or F-theory compactifications at constant (perturbative thus
vanishingly small) coupling, the parity even counterpart of the
GSS counterterm, dictated by supersymmetry, reads \be L_{kin} =
\sum_I v_I C^I_{ab} Tr (F^a F^b) \ee where $v_I$ is an $SO(1,n_T)$
vector and $C^I_{ab}$ is a set of $n_T+1$ structure constants
satisfying \be \eta_{IJ}C^I_{(ab}C^J_{cd)} = \sum_f Tr_{R_f} (T_a
T_b T_c T_d) \ee for anomaly cancellation \ie gauge invariance of
the one-loop effective lagrangian. It is clear that a combination
satisfying \be \eta_{IJ}C^I_{(ab}C^J_{cd)} = 0 \ee is gauge
invariant {\it per se} and is thus not related to one-loop anomaly
cancellation and can always be present even in the absence of
chiral fermions. Notice that contrary to heterotic models the Type
I dilaton lies in a hypermultiplet and does not play a role in
this context \cite{BPSSW, MBJFMinst}. In fact we have already
mentioned that it is possible to construct Type I and F-theory
models with $n_T = 0$ \cite{ABPSS} whose (non-perturbative)
heterotic dual would exist only at a fixed value for the dilaton.
As already observed, the tensor scalars moduli space is \be
SO(1,n_T)/ SO(n_T) \ee

A large class of tractable models is given by F-theory
compactifications on VB orbifolds. These are elliptically fibered
CY threefolds with a base of the form $B= K3/\sigma$ with $\sigma$
an antiholomorphic involution of $K3$ that reverses the
holomorphic 2-form $\sigma \omega_{2,0} = - \omega_{2,0}$. The
resulting CY is given by $X= K3\times T^2/\sigma'$ where $\sigma'$
combines $\sigma$ with the $Z_2$ action $Z\rightarrow - Z$ on the
torus coordinate. As a result the holomorphic 3-form $\omega_{3,0}
= \omega_{2,0} \wedge dZ$ is invariant. The classification due to
Nikulin is given in terms of three integers $(r,a, \delta)$ with
$\delta = 0,2$ representing the `parity' of the canonical class,
$1\le r \le 20$ the rank of the $\sigma$-invariant sublattice of
$H^2(K3,Z)$ and $1\le a \le 11$ the rank of the Picard lattice of
$K3/\sigma$. For $(r,a) \neq (10,10), (10,8)$, the Hodge numbers
of the base $B= K3/\sigma$ and the threefold $X= K3\times
T^2/\sigma'$ are given by \be h_{11}(B) = r \quad , \quad
h_{11}(X) = 5 + 3r -2a \quad , \quad h_{11}(X) = 65 - 3r -2a \ee
Moreover the elliptic fibration degenerates at $k = (r-a)/2$
rational curves (spheres) $E_i$ and at a curve of genus $g = (22 -
r - a)/2$. The degenerations are all of the $D_4$ type, equivalent
to 4 D7-branes on an $\Omega7^-$-plane, \ie a bound state of
7-branes with no monodromy and thus constant dilaton. The
resulting gauge group is $SO(8)^{k+1}$ with $g$ hypers in the
Adjoint of the $SO(8)$ gauge group associated to the curve of
genus $g$. Notice that for sufficiently high $g$ the latter can
completely Higgs this factor but the remaining $k$ are always
unbroken. It is an easy exercise to compute and factorize the
anomaly polynomial \be I_{FTonVB} = 2 \sum_{i=1}^k (X_i - Y)^2 +
[(r-10) + (a-10)] (X_0 - Y) \ee where \be Y = {1\over 32\pi^2}
trR^2 \quad , \quad X_0 = {1\over 8\pi^2} tr F^2_0 \quad , \quad
X_i = {1\over 8\pi^2} tr F^2_i \ee with 0 labelling the group
associated to the genus $g$ curve. It is also easy to check that
the number of tensors, which is $r$ after inclusion of the
self-dual one in the supergravity multiplet, is always larger than
$k+1$, the number of terms in the reducible anomaly polynomial.
One can expect the GSS mechanism to be at work. Notice that the
case (10,10,0) is special and corresponds to the Enriques
involution which has no fixed points where the torus fibration
could degenerate. The anomaly polynomial is exactly zero (since
$n_T = 9$ and $n_H=12$ and $n_V = 0$) and does not require any
GS-like mechanism. The elliptic threefold is the one considered by
FHSV that has $h_{11}(X) = h_{21}(X)=11$. One might be tempted to
associate the octonionic magic model to the VB orbifold (10,4,0)
with gauge group $SO(8)^4$ of rank $16$. However this cannot work
in $D=6$ since one of the $SO(8)$ factor is singlet out wrt to the
other three. The three adjoint hypers can fully break the former
while the latter three remain unbroken. After compactification to
$D=4$, one can turn on VEV's for the complex scalars in the vector
multiplets and go to the Coulomb phase where the gauge group is
broken to its maximal torus (Cartan) and all the charged hypers
can get a mass. This indeed gives a CY threefold compactification
with $h_{11} = h_{21} = 27$ and the correct number of vector and
hyper multiplets in $D=4$ dimensions. Yet it is difficult to
envisage a restoration of a full symmetry among the 16 Cartan
vectors. We believe the correct 6-D description of the Octonionic
magic model requires a different construction in terms of Type I
to which we now turn.

\subsection{The Octonionic Model}

A possible candidate for a 6-D parent of the Octonionic magic
model is a Type I compactification on $T^4/Z_2$ with a peculiar
unoriented projection. In the untwisted sector one can combine
$\Omega$ with an order two shift \`a la Scherk-Schwarz in any of
the internal coordinates \cite{DP, GJ, ABPSS, BZ, ZK}. Although
the massless spectrum, carrying zero KK momentum, is completely
unaffected and gives rise to $\cN=(1,0)$ supergravity coupled to
one tensor and 4 neutral hyper multiplets, the transverse channel
amplitude exposing massless RR tadpole gets crucially modified in
that no massless RR tadpole associated to $\Omega9$-planes is
present. The correct interpretation, possibly after T-duality, is
that one is superimposing an equal number of two mutually
supersymmetric $\Omega$-planes with opposite R-R charge
\cite{BPStor, MBtor, Wtor}. Since the geometric SS shift does
nothing to the winding states that get projected by the Klein
bottle one has still 16 $\Omega5$-planes that carry non-vanishing
RR charge. In the twisted sector this exotic $\Omega$ projection
keeps 8 tensor multiplets and 8 hypermultiplets. In all one has
$n_T^{cl} = 9$ tensor and $n_H^{cl}=12$ neutral hyper multiplets.
Although the field content is non anomalous there is still an
untwisted RR tadpole (not associated to chiral anomalies
\cite{MBJFM, IRU}) to be cancelled. It requires the introduction
of 16 dynamical D5-brane ad their unoriented open string
excitations. The absence of twisted R-R tadpoles, consequent to
the choice of splitting 16 fixed points into 8 (hypers) and 8
(tensors), implies that the Chan-Paton embedding of the $Z_2$
should be freely acting and leads to a $U(16)$ group coupled to
one hypermultiplet in the Adjoint representation. Further details
can be found in the Appendix. The anomaly polynomial is once again
exactly zero and one can go to the Coulomb branch where
$U(16)\rightarrow U(1)^{16}$. The amusing feature of this breaking
pattern is that the 16 vectors are all on the same footing as
required for their being part of an irreducible representation
such as the spinor of $SO(1,9)$. In the Coulomb branch the gauge
couplings are given by \be v_I C^I_{ab} \ee where $C^I_{ab}$ are
the symmetric $\gamma$ matrices of $SO(1,9)$ that satisfy the
cocycle condition as required by gauge invariance. It is amusing
to observe that precisely the same cocycle condition allows a
Fierz rearrangement that is necessary in order to prove
supersymmetry of the vector multiplet Lagrangian in $D=10$!

\subsection{The Enriques FHSV Model}

The Enriques FHSV model can be constructed similarly. One starts,
for instance, with a Type IIB compactification on $T^4/Z_2$ and
performs a Klein bottle projection that combines $\Omega$ with a
$Z_2$ involution of $T^4/Z_2$ without fixed points \cite{DP, GJ,
ABPSS, BZ, ZK, BPStor, MBtor, Wtor}. This is nothing but the
Enriques involution at a sublocus of the moduli space where $K3
\approx T^4/Z_2$.

Contrary to the previous case the resulting unoriented closed
string model is not only anomaly free, in the sense that the
anomaly polynomial exactly vanishes, but also free from R-R
tadpoles in the transverse channel. This prevents the possibility
of introducing D-brane and their open string excitations
altogether. As a consequence the Type I model has $n_T^{cl} = 9$,
$n_H^{cl} = 12$ and $n_V = 0$, that is what is needed to produce
the FHSV model after compactification on a $T^2$.

Alternatively one can construct an equivalent model as an
asymmetric orbifolds of Type IIB. Indeed S-duality of Type IIB in
$D=10$ relates the symmetry $\Omega$ (worldsheet parity) to
$(-)^{F_L}$ (change of sign of all R-R fields). Although
quotienting (\ie `gauging') $\Omega$ and $(-)^{F_L}$ gives
different results in $D=10$, \ie Type I in the former case and
Type IIA in the latter, combining $\Omega$ and $(-)^{F_L}$ with an
order two involution of a compactification leads to equivalent
models in lower dimensions \cite{VWdual, ASdual, CDMP}. For our
purposes one can check that quotienting Type IIB on $K3$ by
$(-)^{F_L}\sigma_\cE$ yields an anomaly free $\cN = (1,0)$ model
with $n_T = 9$, $n_H = 12$ and $n_V = 0$. In particular one can
perform the analysis at a point in the $K3$ moduli space where $K3
\approx T^4/Z_2$ such as in fermionic constructions \cite{ABK, AB,
KLT, FK, FKP, MBthesis} or in Gepner models \cite{ABPSS, Blume,
ABSS}.

\subsection{Other Magic Models}

By using asymmetric orbifolds and free fermion constructions
\cite{ABK, AB, KLT, FK, FKP} Kounnas et al \cite{Hyperfree} have
been able to construct magic hyper-free $\cN =2$ supergravities in
$D=4$. We would like to comment on the possibility of constructing
other $D=6$ models which can play the role of parents for the
magic $\cN =2$ supergravities with $n_V = 8 + 5 + 2 =15$, $n_V = 4
+ 3 + 2 = 9$ and $n_V = 2 + 2 + 2 = 6$, that enjoy $SO(1,5)$,
$SO(1,3)$ and $SO(1,2)$ symmetry respectively since the $D=6$
vector multiplets in the Coulomb phase (after Higgsing)  transform
as spinors of dimension 8, 4 and 2 respectively. Once again it is
amusing to observe that these are precisely the dimensions and
spinor representations that allow consistent supersymmetric
Yang-Mills Lagrangian. The cocycle conditions on the structure
constants that determine the coupling of the scalars in tensor
multiplets to the vector fields are reinterpreted as the
possibility of performing a the necessary Fierz rearrangement on
four Fermi terms that appear after varying the gauge fields.

Many $\cN = (1,0)$ superstring models with $n_T = 5$ and $n_H^{cl}
=16$ are known \cite{MBAS1, MBAS2, ABPSS} with rank higher than 8.
For our purposes, a particularly interesting class are models
where charged hypers transform in the 28-dimensional adjoint of
$SO(8)^2$ or the 28-dimensional of $U(8)$ or the 27-dimensional of
$Sp(8)^2$. The pattern of symmetry breaking in all these cases
yields $U(1)^8$ with neutral hypers. Gravitational anomaly
cancellation fixes the number of neutral hypers once the number of
tensor and vector multiplets is fixed. The former by the choice of
unoriented closed string projection and the latter by the choice
of gauge symmetry breaking pattern which is tantamount to the
choice of Wilson lines on D9's and position of D5's. The models
labelled by $D_{16}, A_{64}$ in \cite{ABPSS} can accomplish the
task. Also F-theory on the VB orbifold (6,4,0) with $SO(8)^2$
gauge group and $g=6$ hypers in the (1, 28) representation could
do the job after compactification to $D=4$. However, as for the
(10,4,0) case, it is hard to envisage the origin of the $SO(1,5)$
symmetry among the vectors in $D=6$.

Fewer models with $\cN = (1,0)$ superstring models with $n_T = 3$
or $n_T = 2$ are known. In order to get $n_V=4$ or $n_V=2$ neutral
vector multiplets coupled to neutral hypers one has to start with
models with at least $U(4)$ or $SO(8)$ or $Sp(8)$ for the former
or $SU(2)$ (which is GS cancellable, lacking a quartic Casimir!).
There are choices that however do not seem to yield the desired
pattern of symmetry breaking. Once again F-theory on VB orbifolds
with $r=4$ and $r=3$ respectively and $a=4$ or $a=2$ and $a=1$
respectively could do the job after compactification on $T^2$ but
obscure the origin of the $SO(1, n_T)$ symmetry among the massless
vectors in the Coulomb phase in $D=6$. For $n_T = 2$, one can
perform a different unoriented projection of the unique Type I
model with $n_T=0$ in $D=6$, based on the $(k=1)^6$ Gepner model
\cite{ABPSS}, and keep $n_T=2$. Stringent constraints from tadpole
cancellation seem however to naively prevent this possibility.

It is not clear that magic supergravity models in different
dimensions have a unique embedding  in superstring constructions.

\section{Further Comments and Conclusions}

Our analysis so far has been essentially classical. Quantum
corrections may {\it a priori} spoil the beautiful geometry of the
two magic models under consideration. However it has been known
for a while that perturbative and non-perturbative corrections to
the 2-derivative effective action vanish in the FHSV Enriques
model \cite{FHSV}. The argument is based on heterotic / Type II
duality \cite{Aspin}. The hypermultiplet geometry is exact in the
heterotic description since the dilaton belongs in a vector
multiplet. The special geometry is exact in the type IIB
description, since the dilaton belongs in a hypermultiplet and no
worldsheet instantons are present since the Enriques CY threefold
is self-mirror. The same sort of argument applies to the
quaternionic magic model. As we will momentarily observe, the
moduli space is a fibration over the moduli space of the FHSV
Enriques model which is uncorrected as we have just seen. Moreover
the massless open string spectrum, consisting in the Coulomb phase
of 16 neutral vector multiplets  and as many hyper multiplets,
enjoys $\cN=4$ supersymmetry and has thus zero $\beta$-function
and produces no corrections to the two-derivative effective
action. Yet there may be interesting threshold corrections to four
and higher derivative terms in the effective action such as the
ones computed in \cite{HM} for the FHSV Enriques model. For
related work on BPS states in the FHSV model see \cite{KM,
JustDavid}.

Before concluding, we would like to comment on the two possible
Higgs mechanisms mentioned in the paper. Notice that a long vector
multiplet (16 states: 8 bosons and 8 fermions) in a 4D sense
corresponds to nonzero VEV for hyper-scalars and zero VEV for
vector-scalars. A short vector multiplet instead (8 states: 4
bosons and 4 fermions) corresponds to zero VEV for hyper-scalars
and non-zero VEV for vector-scalars. Obviously only the former
admits a 6D uplift since there are no BPS particle (point-like)
states in 6D $\cN = (1,0)$ supersymmetric theories.

We would also like to comment on the decomposition of the magic
moduli spaces as fibrations \be M_q = B_q + F_q \ee In (Type I)
string theory the base $B_q$ should describe closed string moduli,
while the fiber $F_q$ describes open string moduli. It is amusing
to observe that the fiber precisely matches (at least for $D=4,5$)
the moduli space of non-BPS attractor solutions \cite{FM}. In all
there are 12 models forming three sequences of four exceptional
geometries, associated to the four division algebras $J_3^R,
J_3^C,J_3^H,J_3^O$. They correspond to $D=5,4,3$ dimensions and
$q=1,2,4,8$, one has \be dimM_q = 3q+(7-D) \quad , \quad dimB_q =
q+(7-D) \quad , \quad dimM_q = 2q \quad , \ee where, depending on
$D$, the dimensions are taken over real (R), complex (C) and
quaternions (H), respectively.

The $D=4$ (special geometries) and $D=3$ (quaternionic geometries)
cases are related to one another by c-map \cite{CFG}. The
decompositions are summarized in the following tables.

{\bf I Sequence ($D=5$)}
 \bea \left.
\begin{array}{|l|l|l|l|}
\hline
 q & {\rm Scalar  \, Manifold} \: M_q & {\rm Base} \: B_q & {\rm Fiber} \: F_q \\
 \hline
  8 & {E_{6(-26)}\over F_4} & {SO(9,1)\over SO(9)}\times {\tiny SO(1,1)} & {F_{4(-20)}\over SO(9)} \\
\hline
 4 & {SU^*(6)\over Usp(6)} & {SO(5,1)\over SO(5)}\times {\tiny SO(1,1)} & {Usp(4,2)\over Usp(4)\times Usp(2)} \\
\hline
 2 & {SL(3,C)\over SO(3)} & {SO(3,1)\over SO(3)}\times {\tiny SO(1,1)} & {SU(2,1)\over SU(2)\times U(1)} \\
\hline
 1 & {SL(3,R)\over SO(3)} & {SO(2,1)\over SO(2)}\times {\tiny SO(1,1)} & {SL(2,R)\over SO(2)} \\
\hline
\end{array}
\right. \nn
 \eea

{\bf II Sequence ($D=4$)}
 \bea \left.
\begin{array}{|l|l|l|l|}
\hline
 q & {\rm Scalar \, Manifold} \: M_q & {\rm Base} \: B_q & {\rm Fiber} \: F_q \\
 \hline
  8 & {E_{7(-25)}\over E_6 \times U(1)} & {SO(10,2)\over SO(10)\times SO(2)} \times {SU(1,1)\over U(1)}
  & {E_{6(-14)}\over SO(10)\times U(1)} \\
\hline
 4 & {SO^*(12)\over U(6)} &  {SO(6,2)\over SO(6)\times SO(2)} \times {SU(1,1)\over U(1)}&
 {SU(4,2)\over SU(4)\times SU(2)\times U(1)} \\
\hline
 2 & {SU(3,3)\over SU(3)\times SU(3)\times U(1)} &  {SU(2,2)\over SU(2)\times SU(2) \times U(1)}
  \times {SU(1,1)\over U(1)} &
 {SU(2,1)\over U(2)}\times {SU(1,2)\over U(2)} \\
\hline
 1 & {Sp(6,R)\over U(3)} &  {Sp(4,R) \over U(2)} \times {SU(1,1)\over U(1)} & {SU(2,1)\over U(2)} \\
\hline
\end{array}
\right. \nn
 \eea

{\bf III Sequence ($D=3$)}
 \bea \left.
\begin{array}{|l|l|l|l|}
\hline
 q & {\rm Scalar \,  Manifold} \: M_q & {\rm Base} \: B_q & {\rm Fiber} \: F_q \\
 \hline
  8 & {E_{8(-24)}\over E_7 \times SU(2)} & {SO(12,4)\over SO(12)\times SO(4)} &
  {E_{7(-5)}\over SO(12)\times SU(2)} \\
\hline
 4 & {E_{7(-5)}\over SO(12) \times SU(2) \times U(1)} & {SO(8,4)\over SO(8)\times SO(4)}
 & {SO(8,4)\over SO(8)\times SO(4)} \\
\hline
 2 & {E_{6(+2)}\over SU(6)\times SU(2)} & {SO(6,4)\over SO(6)\times SO(4)} & {SU(4,2)\over SU(4)\times SU(2)\times U(1)} \\
\hline
 1 & {F_{4 (+4)}\over Usp(6)\times Usp(2)} & {SO(5,4)\over SO(5)\times SO(4)} & {Usp(4,2)\over Usp(4) \times Usp(2)} \\
\hline
\end{array}
\right. \nn
 \eea

Note that the third column of Sequence II has also been recently
found in the framework which relates Magic Models to constrained
instantons \cite{Dasgupta:2007fm}, while the group $E_{8(-24)}$
(first entry in Sequence III) is the exceptional group used in
\cite{Lisi:2007gv} in a (hopeless) attempt to unify gravity with
the Standard Model.

Finally, we would like to comment on the `hyper-free' magic models
of Kounnas, Dolivet and Julia \cite{Hyperfree}\footnote{We would
like to thank B.~Julia and C.~Kounnas for explaining to us their
construction prior to publication.} based on left-right asymmetric
constructions (shift orbifolds or free fermions) with $\cN =
(4,1)$ worldsheet susy \cite{FK, ABK, KLT}. Their construction
consists in a two-step procedure. The first step yields a model
with $\cN=2+4,2+2,2+1$ spacetime susy. The second step breaks all
susy associated to right-movers and yields $\cN=2+0$ spacetime
susy. Differently from `standard' compactifications with $\cN=1+1$
spacetime susy, such as CY compactifications, the axio-dilaton
belongs in a vector multiplet, like in the heterotic string on
$K3\times T^2$, not in the `universal' hypermultiplet! The minimal
hyper-free theory has a single minimally coupled vector multiplet
$S$ ($K = -\log (S + \bar{S})$) associated to the axio-dilaton.

The first non-minimal magic hyper-free theory, associated to the
Jordan algebra $J_3^C$, has 9 $\cN=2$ vector multiplets (18 real
scalars) plus one graviphoton and moduli space
$$
\cM_3 = {SU(3,3)\over SU(3)\times SU(3)\times U(1)}
$$
like in $\cN = 3$ supergravity with 3 vector multiplets, which
therefore are in different representations of the duality group
$SU(3,3)$, the threefold selfdual antisymmetric (for $\cN=2$) and
the fundamental (for $\cN=3$).

The second, associated to the Jordan algebra $J_3^H$, contains 15
vector multiplets (30 real scalars) plus one graviphoton. The
moduli space is
$$
\cM_6 = {SO^*(12)\over U(6)}
$$
like in $\cN = 6$ supergravity with 15 + 1 graviphotons, with
identical transformation properties (32-dimensional real chiral
spinor, after including the magnetic duals) under the duality
group $Spin^*(12)$\footnote{This can be taken as evidence that
supersymmetric completions of theories with the same bosonic
sector may differ from one another. Indeed $\cN=2$ and $\cN=6$
supergravities differ even at the level of the fermionic
spectrum.}.

\section*{Acknowledgements}

One of us (S.~F.) would like to acknowledge illuminating
conversations with R.~Varadarajan on fibrations. The work of M.~B.
has been supported in part by the European Community Human
Potential Program under contract MRTN-CT-2004-512194, by the INTAS
grant 03-516346, by MIUR-COFIN 2003-023852, and by NATO
PST.CLG.978785. The work of S.~F. has been supported in part by
the European Community Human Potential Program under contract
MRTN-CT-2004-005104 "Constituents, fundamental forces and
symmetries of the universe" and under contract MRTN-CT-2004-503369
``The quest for unification: Theory Confronts Experiments'', in
association with INFN Frascati National Laboratories, by INTAS
under contract 05-7928, and by D.O.E. grant DEFG03- 91ER40662,
Task C.

\newpage
\section*{Appendix: Parent Type I models}

In this appendix, we describe the one-loop partition functions
encoding the spectra of the Type I models in $D=6$ that give rise
to the two magic supergravity models in $D=4$ after
compactification on $T^2$. In both cases, one starts from Type IIB
on $T^4/Z_2\approx K3$.  In the untwisted sector, one
has\footnote{For notational simplicity we omit the (regulated)
contribution of the non-compact bosonic zero-modes and the modular
integration measure.} \be \cT_u = {1\over 2} \left[ \vert \sum_\a
c_\a {\theta^4_\a \over \eta^{12}} \vert^2 \Lambda_{(4,4)} + 16
\vert \sum_\a c_\a {\theta^2_\a \theta_\a^2({1\over 2}) \over
\eta^{6}\theta_1^2({1\over 2})} \vert^2 \right]\ee where
$\theta_\a$ are Jacobi functions, $\eta$ is Dedekind function,
$\Lambda_{(4,4)}$ denotes the sum over generalized momenta
$\vec{p}_{L/R} \approx \vec{p} \pm \vec{w}$ and $c_\a$ enforce the
GSO projection. The massless spectrum consists in the $\cN =
(2,0)$ supergravity coupled to 5 tensor multiplets. In the twisted
sector, one has \be \cT_u = {16\over 2} \left[ \vert \sum_\a c_\a
{\theta^2_\a \theta_\a^2 ({\tau\over 2})\over
\eta^{6}\theta_1^2({\tau\over 2})} \vert^2 + \vert \sum_\a c_\a
{\theta^2_\a \theta_\a^2({1 + \tau\over 2}) \over
\eta^{6}\theta_1^2({1+\tau\over 2})} \vert^2 \right]\ee Each of
the 16 terms produces one massless $\cN = (2,0)$ tensor multiplet.
In all one thus has 21 $\cN = (2,0)$ tensor multiplets. Each one
of them decomposes into one $\cN = (1,0)$ tensor- and one $\cN =
(1,0)$ hyper-multiplet.

For the parent of the quaternionic magic model, the Klein bottle
(unoriented) projection in the untwisted sector combines
world-sheet parity $\Omega$ with an order two shift
$\sigma_{\vec\delta}$ in the internal directions \be \cK_u =
{1\over 2} [P_{\vec\delta} + W_{\vec{0}}] \sum_\a c_\a
{\theta^4_\a \over \eta^{12}}  \ee where $P_{\vec\delta}$ denotes
the projected sum over momenta, while $W_{\vec{0}}$ denotes the
(unprojected) sum over windings. As a result, the unoriented
closed string spectrum at the massless level consists in $\cN =
(1,0)$ supergravity coupled to one tensor multiplet and four
neutral hypermultiplets. In the twisted sector, one has \be \cK_t
= {8-8 \over 2} \sum_\a c_\a {\theta^2_\a \theta_\a^2 ({\tau\over
2})\over \eta^{6}\theta_1^2({\tau\over 2})} \ee that yields 8
tensor multiplets and as many hypermultiplets. In all one has
$n_T^{cl} = 9$ and $n_H^{cl} = 12$. In the transverse channel, the
only massless RR tadpole comes from $\tilde{P}_{\vec{0}}$
generated by the modular $S$ transformation of term with
$W_{\vec{0}}$. In order to cancel the $\Omega 5$ tadpole, one has
to introduce $N=16$ D5-branes and their images. In the transverse
channel, the Annulus and M\"obius-strip amplitudes read \be
\tilde\cA_{55} = {1\over 2\cdot 32 } [ 2 \tilde{P}_{\vec{0}}
N\bar{N} + \tilde{P}_{\vec{\delta}} N^2 +
\tilde{P}_{\vec{\delta}}\bar{N}] \sum_\a c_\a {\theta^4_\a \over
\eta^{12}}  \ee \be \tilde\cM_{5\Omega} = - {2\over 2}
[\tilde{P}_{\vec{\delta}} N + \tilde{P}_{-\vec{\delta}}\bar{N}]
\sum_\a c_\a {\theta^4_\a \over \eta^{12}}  \ee The resulting
massless spectrum consists in vector and hyper multiplets in the
adjoint representation of $U(16)$. Spontaneous symmetry breaking,
which is equivalent to moving the D5 branes, produces
$U(16)\rightarrow U(1)^{16}$.

For the parent of the Enriques FHSV model, one starts with the
same Type IIB compactification on $T^4/Z_2$ as in the previous
case. In the untwisted sector, the Klein bottle projection
combines world-sheet parity with an order four rotation,
equivalent two an order two projection $\vec\delta$ on both
windings and momenta \be \cK_u = {1\over 2} [P_{\vec\delta} +
W_{\vec{\delta}}] \sum_\a c_\a {\theta^4_\a \over \eta^{12}} \quad
. \ee This has no effect on the massless states so that the
untwisted unoriented closed string spectrum consists in $\cN =
(1,0)$ supergravity coupled to one tensor multiplet and four
neutral hypermultiplets. In the twisted sector, one has the same
Klein bottle projection $\cK_t$ as above, that yields 8 tensor
multiplets and as many hypermultiplets. In all one has $n_T^{cl} =
9$ and $n_H^{cl} = 12$. In the transverse channel $\cK \rightarrow
\tilde{\cK}$ produces non massless RR tadpoles at all. As  a
consequence neither D9- nor D5-branes can be introduced and the
model is a consistent unoriented closed string theory without open
strings.

\end{document}